# Nuclear spin pair coherence in diamond for atomic scale magnetometry

Nan Zhao, Jian-Liang Hu, Sai-Wah Ho, Tsz-Kai Wen & R. B. Liu

*Department of Physics, The Chinese University of Hong Kong, Shatin, N. T., Hong Kong, China*

**The nitrogen-vacancy (NV) centre, as a promising candidate solid state system of quantum information processing[1-3], its electron spin coherence is influenced by the magnetic field fluctuations due to the local environment. In pure diamonds, the environment consists of hundreds of $^{13}$C nuclear spins randomly spreading in several nanometers range forming a spin bath[4-8]. Controlling and prolonging the electron spin coherence under the influence of spin bath are challenging tasks for the quantum information processing. On the other hand, for a given bath distribution, many of its characters are encoded in the electron spin coherence. So it is natural to ask the question: is it possible to 'decode' the electron spin coherence, and extract the information about the bath structures? Here we show that, among hundreds of $^{13}$C bath spins, there exist strong coupling clusters, which give rise to the millisecond oscillations of the electron spin coherence. By analyzing these oscillation features, the key properties of the coherent nuclear spin clusters, such as positions, orientations, and coupling strengths, could be uniquely identified. This addressability of the few-nuclear-spin cluster extends the feasibility of using the nuclear spins in diamond as qubits in quantum computing. Furthermore, it provides a novel prototype of single-electron spin based, high-resolution and ultra-sensitive detector for the chemical and biological applications.**

As is well known, nuclear magnetic resonance (NMR) techniques measure the nuclear spin coherence signals, and translate these signals to the information of the local environment of the nuclear spins, which help us determined the properties of the host samples[9], such as the structures, interactions and dynamics. Due to the small magnetic moment each nuclear spin carries, the NMR signals always have to be obtained by the ensemble measurements. This low sensitivity limits its application in small size systems, such as in the nanoscale detection. To enhance the sensitivity, electron spins would have more advantages, since their magnetic moments are about $10^3$ times larger than those of nuclear spins. By the optical detected magnetic resonance (ODMR) technique[10], single electron spin in diamond was used in nanoscale sensing and imaging of weak magnetic field distributions of the detected samples[11-14]. In this paper, taking into account the quantum nature of both detector and detected spins, we propose a detector based on the quantum entanglement of single electron spin in diamond and the detected nuclear spin clusters. Our proposal enables a detector with few-nucleus sensitivity and sub-nanometer resolution under room temperatures.

The negatively charged NV centre traps electrons and forms a spin triplet for its ground state. Its excited level structures enable an optical polarization of the ground state spin through the inter system cross[15-18]. The polarized spin state serves as the initial state of the subsequent electron spin resonance (ESR) operations by microwave (MW) pulses. After MW operations, the spin state is read out by measuring the photoluminescence (Fig.1a). The ODMR of single NV centre provides a key to investigate the local spin environment around it. In type Ib diamond, the NV centre electron spin interacts mainly with the doped nitrogen electron spins. The spin coherence of NV centre exhibits different behaviors under different bath properties tuned by magnetic fields and temperatures[19, 20]. In the ultra-pure type IIa diamond samples, which we will focus on, the spin coherence time is measured as several hundreds of micro seconds in the room temperature[21], and decoherence is believed to be dominated by the coupling to the surrounding $^{13}$C nuclear spins[4-8].

Each $^{13}$C nuclear spin in the bath does not play the equal role to the electron spin coherence depending on its distance from the NV centre. A few proximal nuclear spins, which coherently coupled to the electron spin, is responsible to the fast oscillations in the coherence signal[4]. While the remainder spins with large distance and thus weak couplings, provides a relative smooth background decay of the electron spin coherence. The coherent coupled nuclear spins are shown to be successfully addressed and used as a quantum registers[5, 6]. However, for the natural abundance $^{13}$C samples, the low possibility of the $^{13}$C appearing in the proximal sites around the NV centre limits the resources and the applications.

Similar to the single nuclear spin dynamics, the spin clusters, which contain several interacting nuclear spins, could also be classified as coherent or incoherent ones according to the intra-bath dipolar interactions and the hyperfine interactions to the electron spin. The interactions between the bath spins in the coherent clusters are strong enough that the nuclear spins undergo several flip-flops during the electron coherence time. While for the bath spins in the incoherent clusters, they evolve slowly due to the weak interactions and induce the irreversible decay of the electron spin coherence. Oscillations of electron spin coherence due to the coherent nuclear spin cluster were noticed in recent experiments and theoretical simulations1[4, 22]. These observations motivate us to address this question: is it possible to extract the information about the coherent cluster from the electron spin coherence oscillations? In the following, we show that, analyzing the distinguished oscillation features, strong coupled $^{13}$C nuclear pairs could be precisely identified.

The simplest coherent cluster is a $^{13}$C pair with the two nuclear spins located at a nearest neighbor lattice site (Fig. 1a). Although the probability of two $^{13}$C atom occupying the same carbon bond is $p^2 \sim 10^{-4}$ for the sample with natural abundance $p = 1.1\%$, Fig. 1c shows that it has a probability of 50% to find such a coherent pair in the ~1.5 nm range around the NV centre. These two nuclear spins are coupled through the dipolar interactions with the typical strength ~2kHz. At the same time, the hyperfine interactions for the nuclear spin with ~1nm from the NV centre are also the order of kHz (Fig. 1d). As a result, in order to fully exhibit the nuclear spins pair dynamics, it is important to prolong the coherence time $T_2$ to millisecond or longer.

In our previous investigations, we have shown that applying a magnetic field larger than 50G, could suppress the environment magnetic field fluctuations arising from the single nuclear spin rotations, and the coherence time is thus extends by a factor of 10, to the order of 100 microseconds. In addition, the dynamical decoupling control pulse sequence is also shown to be an efficient method to preserving the electron spin coherence in the nuclear spin bath[23, 24]. In Fig. 1e-1i, we show the electron spin coherence under the optimal dynamical decoupling[25] for a randomly generated $^{13}$C configuration with a coherent pair located 1.2nm from the NV centre. Under the five-pulse Uhrig dynamical decoupling sequence (UDD5), the spin coherence is prolonged to ~2ms, and the coherent modulations of the spin coherence at millisecond scale are clearly shown. The contribution of the coherent nuclear spin pair to the electron spin coherence is singled out in Fig. 1e-1i. Its oscillations match the total spin coherence features quit well.

The physical picture of the coherent pair oscillations is understood with the help of the pseudo-spin model shown in Fig. 1b. Under an applied magnetic field, which suppresses the energy non-conserving non-secular flipping of the nuclear spins, the two-spin anti-parallel states $|\uparrow\downarrow\rangle$ and $|\downarrow\uparrow\rangle$, can be mapped to a pseudo-spin. The secular parts of dipole-dipole interactions induce the pair flip-flop of the pseudo-spin, i.e. pseudo-field $b_x$. Meanwhile, the coherent pair is subject to the hyperfine field produced by the magnetic moment of electron spin:

$$b_i = \frac{\mu_0}{4\pi} \frac{\gamma_e}{r_i^3} \left[ S^z - 3n_i^z (S \cdot n_i) \right]$$

This effective field difference felt by the two spins provides the effective energy bias of the two pseudo-spin states, i.e. the pseudo-field $b_z = b_1 - b_2$. Under these pseudo-fields, the coherent pair contribution to the full coherence is calculated by:

$$L(t) = \mathrm{Tr}\left[ e^{it_1 \mathbf{b}_0 \cdot \boldsymbol{\sigma}/2} e^{it_2 \mathbf{b}_1 \cdot \boldsymbol{\sigma}/2} \cdots e^{-it_n \mathbf{b}_{s'} \cdot \boldsymbol{\sigma}/2} e^{-it_n \mathbf{b}_s \cdot \boldsymbol{\sigma}/2} \cdots e^{-it_2 \mathbf{b}_0 \cdot \boldsymbol{\sigma}/2} e^{-it_1 \mathbf{b}_1 \cdot \boldsymbol{\sigma}/2} \right]$$

With this pseudo-spin model, one finds that the modulation features of the electron spin coherence, such as the positions and heights of the dips and peaks, are physically determined by the hyperfine field gradient at the position of the coherent pair. We further demonstrate in Fig. 2, this modulation is very sensitive to the change of the hyperfine field felt by the coherent pair. Varying the external magnetic fields, one could tune the direction and magnitude of magnetic moment of the electron spin (Fig. 2a), and thus the hyperfine field felt by the coherent pair. In general, coherent pairs located at different lattice sites will give distinguished modulations to the electron coherence. As an example, we show the calculated electron spin coherence for the four neighboring coherent pairs in the same cubical unit cell with the same orientation (Fig. 2c and d). All the other bath spins are kept the same, and thus provide the same background decay. The differences of the coherence behavior in the four cases attribute to the changes of the pair positions. It is evident that the coherence modulations under various magnetic fields provide a fingerprint feature of the coherent pair. Since the nuclear pair always resides digitally on the lattice sites, the sensitivity and the fingerprint features shown in Fig. 2 enables a precise addressing of the nuclear spin pair in the diamond lattice. This will greatly extends the existing schemes of quantum information processing using nuclear spins.

The precise correspondence of the coherence oscillation features and the one obtained by the pseudo-spin model (see Fig. 2c and d) enables us building a 'fingerprint library', which archives the all information of the coherence modulation due to each possible coherent nuclear spin pairs around the NV centre. By searching the fingerprint library, and comparing with the measured full coherence data, the coherent pair could be addressed. In Fig. 3, we show a typical searching and addressing process for a randomly given $^{13}$C configuration. At the initial stage, all the nearest neighbor carbon pairs (more than 20000 pairs) within the distance of 1.5nm from the NV centre form the possible candidate coherent clusters. For the first round, the electron spin coherence subject to this given spin bath is compared with each pair contributions. Only the pairs give compatible features to the full coherence are retained for the next step comparisons. After this first round filtration, about 100 candidates are kept. Then, the magnetic field is changed, and the second round filtration is performed. In each step, the pairs whose contribution qualitatively conflicts with the given full coherence are filtered out (see lower panels of Fig. 3). This searching strategy ensures the robustness against the inaccuracy and possible error in the coherence measurements. To our experience, within five steps of comparisons and filtrations, the unknown coherent pair could be uniquely identified.

Similar to the single nuclear spin addressing, the addressability of the coherent pair depends on the distance of the coherent pair from the NV centre. For the coherent pair within the 1.5nm, the pair could be precisely addressed. While for larger distances, due to the weaklings of the hyperfine interactions, the coupled $^{13}$C pair will be merged in the noises produced by other spins. In another words, the coupled pair become incoherent for the large distance. However, we point out that in the isotope engineered diamonds[26, 27], the distance could be greatly extended, and NV centre in the $^{12}$C isotope purified diamond could be used as an ultra-sensitive detector in detecting external nuclear spins.

Since the qubit-bath and intra-bath dipolar interactions, which dominate the main decoherence process, are both $r^{-3}$ dependent on the distance, so that the total coherence time due to the spin bath is inversely proportional to the spin concentration. By this scaling relation, reducing the $^{13}$C concentration by a factor $\alpha$, will prolong the total coherence time by the same factor $\alpha$, and the detectable distance of the coherent pair by a factor of $\alpha^{1/3}$. This scaling relation provides an opportunity to not only addressing the intrinsic coherent pair in the diamond itself, but also the ability of detecting the nuclear spin pairs in external molecules.

To demonstrate this point, we consider the configuration depicted in Fig. 4a. An NV centre is planted several nanometers beneath the surface of the $^{12}$C isotope purified diamond. An external molecule to be detected, say $C_{60}$ demonstrated here, is placed on the surface, and could be moved by a tip. The abundance $^{13}$C $p=0.03\%$ is used in the calculations, which implies a scaling factor $\alpha \approx 30$ to the natural abundance. The electron coherences under the influence of the surrounding nuclear spins in two different cases are analyzed. In each case, the $C_{60}$ molecule has two nuclear spins located on different sites of the molecule (insets of Fig. 4a). Besides the dipolar interactions between the two nuclear spins, they are subject to the hyperfine field produced by the NV electron spin, see Fig. 4b. This hyperfine field will change while the molecule is moving of rotating along certain paths. Typical position and angle dependences of coherence signals are shown in Fig. 4c. It is evident that a displacement or rotation in sub-nanometer level will induce significant changes in the electron spin coherence. Analyzing these data, similar to the process of addressing the coherent pair in the diamond, it is possible to extract important information about the few nuclear spins in the single molecule. This detection of few nuclear spins in the single molecule level will be difficult or even impossible through other traditional method.

In summary, a prototype of single electron spin detector in diamond is proposed based on the analyzing of coherent cluster contributions to the spin coherence. The concept of coherent cluster extends the existing understandings of the proximal single nuclear spins strongly coupled to the NV centre. With this concept, the possibility of observing more structures of the spin bath beyond the proximal individual nuclear spins is demonstrated. As an example, coherent nuclear spin pair is analyzed in detail. We predict that the coherent oscillations could be observed by suppressing the unwanted noise and prolonging the coherence time by dynamical decoupling control pulse sequence. The addressing of the coherent pair in the diamond around the NV centre provides new resources for the quantum information processing using the nuclear spin pairs. In the isotope purified diamond, the NV centre could be used as a detector with few-nucleus spin sensitivity and sub-nanometer resolution to detect the structure of external molecules. The ability of detecting the labeled molecule with more complex clusters using this kind of detector will possibly have great application potential in chemistry and biology.

**Figure 1 | Coherent nuclear spin pair in NV centre.  a**, Structure and energy levels of NV centre and coherent nuclear spin pair in diamond.  **b**, Pseudo-spin model. Coherent $^{13}$C nuclear spin pair, which is represented by a pseudo spin on a Bloch sphere, evolves under the pseudo-field $b_x$ and $b_z$ conditioned on the electron spin states. The distance $\delta$ measures the electron spin decoherence due to the nuclear spin pair. **c**, Possibility density and accumulated possibility of the coherent nuclear spin pair as functions of the distance from the NV centre. **d**, A typical Hyper-fine field distribution near the NV centre. **e-j**, Spin coherence (red lines) under dynamical control pulses UDD1~5. The millisecond oscillation features arise from the coherent pair contributions (black lines).

**Figure 2 | Position sensibility and fingerprint features of coherent pairs. a** The central spin states (red, blue and green arrows) under the magnetic field **B** (black arrow) in the direction with angle $\theta$ from [111] direction. **b** central spin energy levels and the corresponding spin components of the $|m=0\rangle$ state. **c** & **d** sensitivity of spin coherence to the nuclear pair positions. The electron spin coherence under four cases, in which the coherent nuclear spin pairs are located at four neighboring sites in the same cubical unit cell (inset of **c**), is compared. The calculated coherence shows fingerprint features of each case (four columns in **d**). Excellent correspondence of the oscillation features between the full coherence signal (lower panels in **d**) and the pair contributions (upper panels in **d**) are shown.

**Figure 3 | A typical process of identifying the coherent pair.** The position of a coherent pair is identified by three-step comparisons of the coherence signal

under different magnetic fields. For given coherence signals (gray curves with shadows), the pair contributions are compared. The pairs which provide compatible oscillation feature are retained for the next-step comparison (typical examples shown upper panels **a**, **c**, and **e**). Otherwise it is filtered out if qualitatively conflicts with the coherence signal (typical examples shown in lower panels **b**, **d**, and **f**). The only one coherent pair, whose oscillation features are compatible with all three coherence signals, is identified after the comparisons (in **e**).

**Figure 4 | Detecting nuclear spins in external molecule. a**, schematic illustration of the detection. A $C_{60}$ molecule with two $^{13}C$ nuclear spins is placed on the surface of a purified diamond with NV centre under the surface. Two cases, in which the two $^{13}C$ nuclear spins are separated by $d_1$ and $d_2$ respectively, are considered. **b**, hyperfine field distribution on the diamond surface. The motion paths (blue dashed lines) of the coherent spin pair are shown for the translations along *y*-axis and rotations about the origin. **c**, position dependence of the coherence signal under the translations along *y*-axis is shown in the left panel. Different features of the coherence patterns in two cases with nuclear spin separation $d_1$ (middle panel) and $d_2$ (right panel) can be used to distinguish the molecule with different nuclear spin distributions.

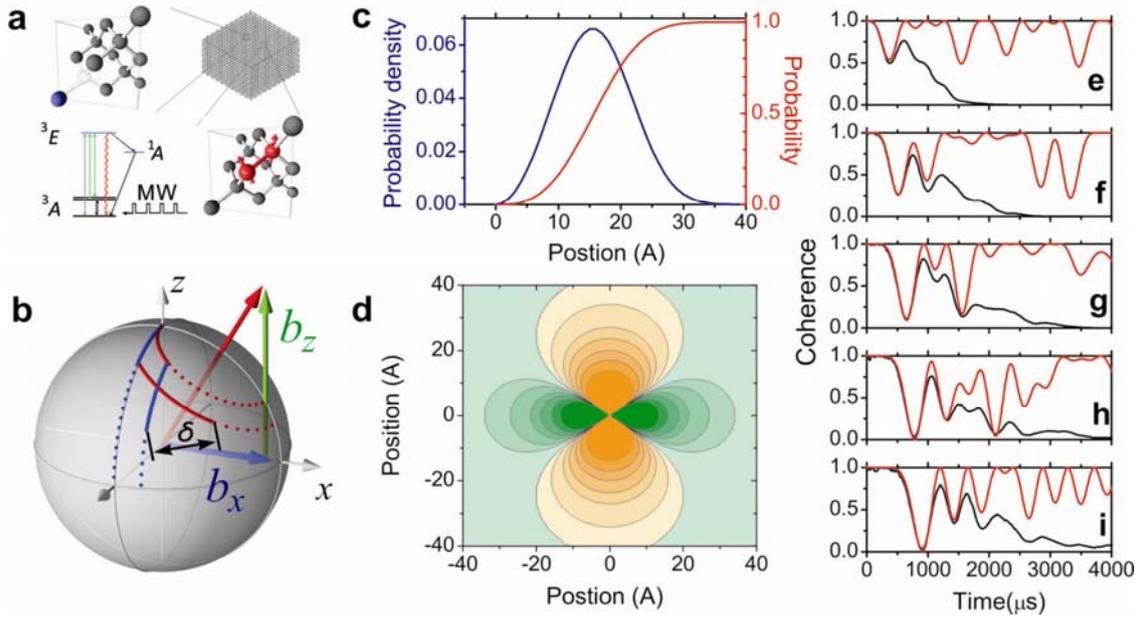

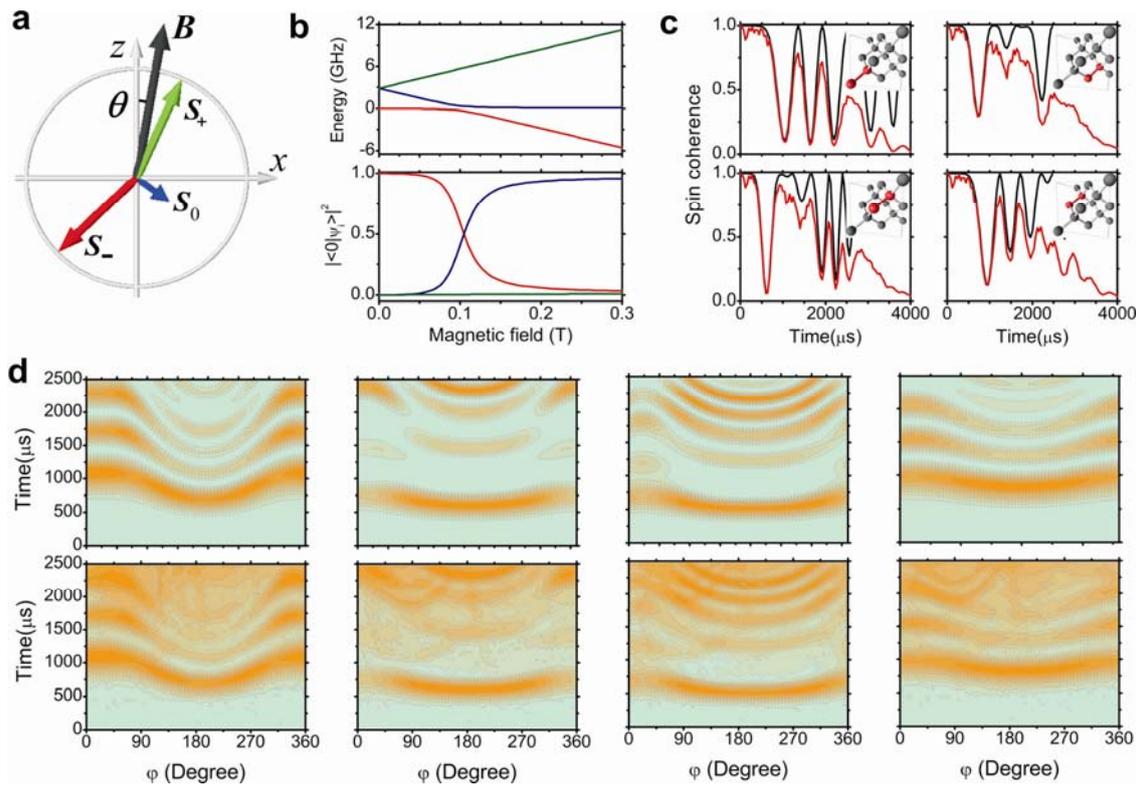

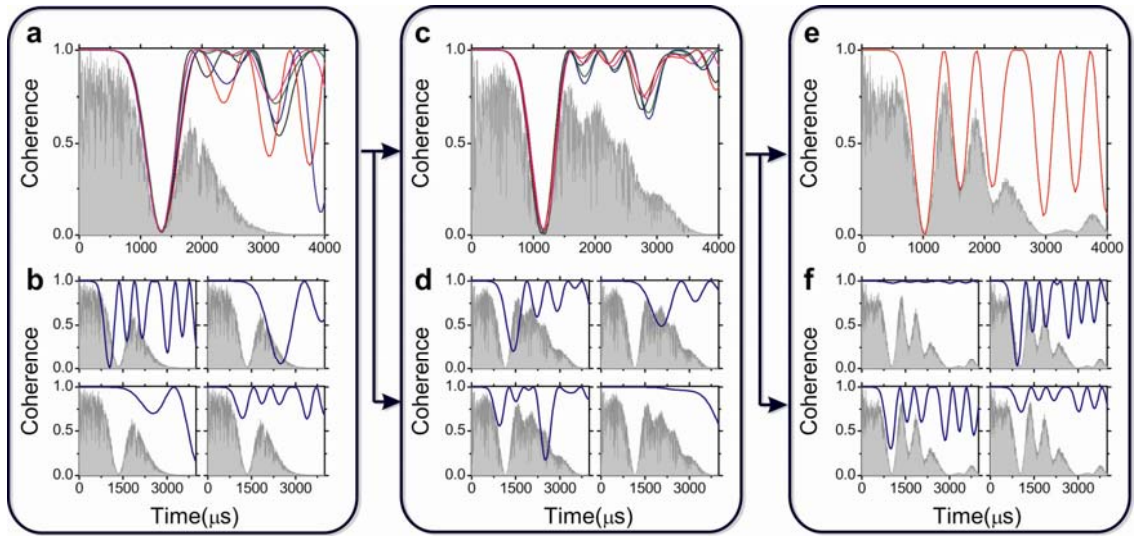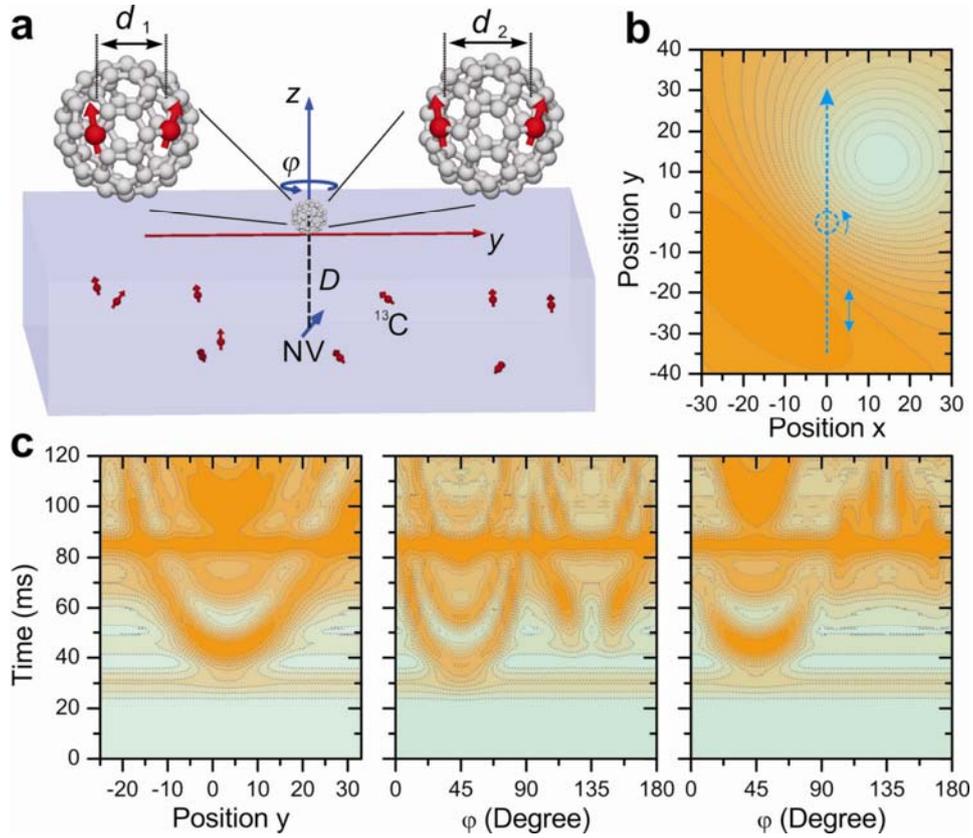